\documentclass[pdflatex,sn-basic]{sn-jnl}
\usepackage{graphicx}%
\usepackage{multirow}%
\usepackage{amsmath,amssymb,amsfonts}%
\usepackage{amsthm}%
\usepackage{mathrsfs}%
\usepackage[title]{appendix}%
\usepackage{xcolor}%
\usepackage{textcomp}%
\usepackage{manyfoot}%
\usepackage{algorithm}%
\usepackage{algorithmicx}%
\usepackage{algpseudocode}%
\usepackage{listings}%
\usepackage{enumitem}

\usepackage{ragged2e}
\usepackage[table]{xcolor} %
\usepackage{caption}
\usepackage{subcaption} 
\usepackage[most]{tcolorbox} 
\usepackage{xltabular} %
\usepackage{ragged2e}
\usepackage{pifont}
\usepackage{booktabs}
\usepackage{rotating}
\usepackage{tabularx}
\usepackage{makecell}
\usepackage[utf8]{inputenc}
\usepackage[dvipsnames]{xcolor}
\usepackage[normalem]{ulem}

\theoremstyle{thmstyleone}%
\theoremstyle{thmstyletwo}%

\theoremstyle{thmstylethree}%

\begin{document}

\title[Test Code Review in the Era of GitHub Actions]{Test Code Review in the Era of GitHub Actions: A Replication Study}








\author[1]{\fnm{Hui} \sur{Sun}}
\author[1]{\fnm{Yinan} \sur{Wu}}
\author[1]{\fnm{Wesley} \sur{Assunção}}
\author[1]{\fnm{Kathryn T.} \sur{Stolee}}
\affil[1]{\orgname{North Carolina State University}, \orgaddress{ \city{Raleigh}, \state{NC}, \country{USA}}}


\abstract{Test code is indispensable in software development, ensuring the correctness of production code and supporting maintainability. Nonetheless, errors or omissions in the test code can conceal production defects. While code review is widely adopted to assess code quality and correctness, little research has examined how test code is reviewed. Spadini et al.'s research on Gerrit (a pre-commit review model) found that test code receives significantly less discussion than production code. However, the most popular review model is currently based on pull requests (PRs), in which contributors propose changes for discussion and approval, a more negotiable and flexible model compared to Gerrit. Furthermore, GitHub Actions (GHA) has become widely used to automate pre-checks and testing, potentially impacting review practices. This leads us to explore whether Spadini et al.'s findings still hold for the PR model in the era of GHA? Our work replicates and extends their work. We focus on GitHub PRs and analyze six open-source projects. We investigate the impact of the PR model and GHA on test code review. Our results show that GitHub’s PR model fosters more balanced discussions between test and production files than Gerrit, albeit with lower overall comment density. However, despite cross-project heterogeneity, GHA adoption triggered a sharp pivot toward production code. Post-GHA, for PRs involving tests, both review probability and comment density reached a median of zero.
These findings reveal how evolving continuous integration pipelines can marginalize test code review. The observed decline in test-centric discussion under GHA warrants concern regarding long-term software quality. Our work also presents recommendations for stakeholders involved in the software development life cycle.
}


\maketitle

\section{Introduction}\label{sec1}

Manually verifying and validating large-scale software systems is both time-consuming and error-prone~\citep{barr2014oracle,myers2004art}. Thus, automated software testing has become indispensable in modern software development~\citep{garousi2016and}. Test code (i.e., code written to verify production code) plays a dual role it: (i) helps establish the correctness of production code, and (ii) serves as executable documentation that supports long-term system maintainability. When integrated into Continuous Integration (CI) pipelines, test cases provide rapid, repeatable feedback that shapes developers’ day-to-day decision-making~\citep{duvall2007continuous}.
However, test code is itself software and is therefore subject to defects, omissions, and design flaws~\citep{van2001refactoring, athanasiou2014test, zaidman2011studying}. Faulty or incomplete tests can mask defects in production code, provide false confidence, or even institutionalize incorrect behavior~\citep{vahabzadeh2015empirical}. As a result, the quality of test code critically depends not only on automated execution, but also on human inspection. This is where code review plays a central role in identifying defects, improving design, and facilitating knowledge transfer~\citep{bacchelli2013expectations, sadowski2018modern, uchoa2021predicting}. 



Many researchers have focused on enhancing the efficiency, quality, and outcomes of code review from a technical perspective, e.g., by analyzing repository data and providing reviewer recommendations~\citep{balachandran2013reducing,tufano2022using,siow2020core}. Also, empirical studies have examined review strategies, influencing factors, and performance metrics, such as determinants of review efficiency and performance ~\citep{armstrong2017broadcast, baysal2013influence, baysal2016investigating}. Yet, relatively little research has explored in detail how test code is reviewed. Reviewing test code poses distinct challenges compared to reviewing production code~\citep{li2019intent, horikawa2025does}: reviewers must reason about test intent, adequacy, and coverage rather than just functional behavior.  Reviewing test code has been consistently emphasized in repository culture and the PR template in recent years~\citep{li2022follow, zhang2022consistent}. Therefore, understanding how developers review test code, and how much attention it receives relative to production code, is essential to assessing the effectiveness of modern code review practices~\citep{morales2015code, kononenko2015investigating}.

Spadini et al.~\citep{spadini2018testing} provided the first examination of how test code is reviewed on Gerrit,\footnote{Gerrit: A platform primarily focused on centralized, mandatory review and linear history, mostly used for enterprise
development of large-scale engineering projects such as Android and OpenStack. \href{https://www.gerritcodereview.com/}{https://www.gerritcodereview.com/}} including the proportion of reviews for test code versus production code, as well as the specific challenges reviewers face when reviewing test code.
However, it remains unclear whether these findings generalize to other review models. While Gerrit follows a pre-commit model with mandatory review requirements, GitHub adopts a PR review model in which contributors propose changes for discussion and approval, which characterizes a more negotiable, flexible review process. Whether review is mandatory on GitHub depends on project settings: some repositories require approval before merging, while others permit self-merging. 
Given this new context, we argue that Spadini et al.'s analyses, conducted eight years ago on the Gerrit platform, warrant re-examination. This is particularly relevant, as GitHub Actions (GHA), introduced in 2019 and rapidly adopted, automates many pre-checks that were previously performed manually by reviewers~\citep{wessel2023github, decan2022use}. To the best of our knowledge, our study is the first to investigate how such automation has influenced test code review practices. 

The goal of this work is to replicate and extend the work by Spadini et al.~\citep{spadini2018testing} (referred to as the \textit{``original study''}). First, our empirical study replicates the \textit{original study}'s quantitative and qualitative analyses using the GitHub PR review model. Furthermore, we extend their research by examining how the introduction of GHA affects reviewer and developer practices regarding test code. Our findings reveal notable differences from the \textit{original study}, reflecting the evolving dynamics of code review models and testing practices in the era of integrated CI tools such as GHA. Our findings can be summarized as follows:

\begin{itemize}
    \item In contrast to Gerrit’s pre-commit model, which enforces mandatory reviews and shows a strong bias toward production code, GitHub’s PR model produces fewer review comments, but distributes reviewer attention more evenly across production and test files. However, with the introduction of GHA, we observe a significant and immediate decline in reviewer attention to test code, as evidenced by the sharp reduction in discussion comments.

    \item Our replication reveals structural differences in how reviewers discuss test code: on GitHub, reviewers focus more on superficial code improvements (35\% to 55\%), while devoting less attention to defect detection and comprehension than on Gerrit. Notably, after the adoption of GHA, discussions on both test improvement and defect detection topics further declined, suggesting a shift of reviewer attention away from the quality of test code.

    \item Our analysis of PRs that change both production and test files further reveals a platform-consistent pattern: consistent with prior interview-based findings, reviewers on both Gerrit and GitHub tend to direct their initial attention to production code. Moreover, our regression analysis shows that reviewers’ initial attention is largely determined by the magnitude of production code churn, rather than by the adoption of GHA.
\end{itemize}

By examining test code review in the context of GitHub's PR model and the adoption of GHA, this study fills a critical hole in the literature on how evolving automation reshapes human review practices. Beyond replication, our results establish an empirical baseline for understanding automation-induced shifts in reviewer attention. This baseline is particularly important as Large Language Models (LLMs) increasingly enter the software development workflow. Without accounting for the deep-seated and measurable influence of existing CI/CD pipelines (e.g., GHA) future studies risk conflating the effects of long-standing automation with those introduced by LLM-based tools. Therefore, our findings caution against attributing changes in review behavior solely to emerging AI technologies, emphasizing the need to disentangle their impact from that of established CI infrastructure.

In summary, the research contributions of our study are as follows: \textbf{I.}~A replication of  Spadini et al.’s analysis of test code review practices in Gerrit, focusing on the PR review model adopted on GitHub. 
\textbf{II.}~Analysis of how the adoption of GHA affects the comment behavior of reviewers on test code, using six diverse open-source projects and combined quantitative and qualitative methods for analysis. 
\textbf{III.} A comparison of our results with those of the \textit{original study}~\citep{spadini2018testing}, showing that the adoption of GHA shifted reviewer attention toward production code. 
\textbf{IV.} Discussion of the potential impacts of the findings on developers, reviewers, tool designers, and researchers. 
\textbf{V.}~A publicly available replication package~\citep{gha_test_review_2026} containing all raw data, scripts, codebook, and analysis results to facilitate reproducibility of our study.

\section{Original Study}\label{sec2}

The study by Spadini et al.~\citep{spadini2018testing} represents the first large-scale empirical investigation explicitly focused on code review practices for test code. Motivated by prior evidence that test code is frequently defective and yet often treated as a second-class artifact~\citep{athanasiou2014test, vahabzadeh2015empirical, zaidman2008mining}, the authors set out to understand whether test files are reviewed, how rigorously they are reviewed, what reviewers discuss, and which challenges developers face when reviewing tests.
To do that, their study adopts a mixed-methods research design. Quantitatively, the authors analyze more than 300,000 code reviews from three large open-source projects (Eclipse, OpenStack, and Qt) hosted on Gerrit~\citep{milanesio2013learning}, a platform dedicated to code review. Qualitatively, they complement this analysis with manual content analysis of review comments and semi-structured interviews with 12 developers from both open-source and industrial contexts.

A key result of the study is that test files are no less defect-prone than production files, suggesting that test code deserves the same level of review as production code. Despite this, the authors find that test files are discussed significantly less often than production files, particularly when test and production code are reviewed together in the same change. When test-only reviews occur, however, test files receive comparable review attention to when they are reviewed alongside production files.
Through manual classification of review comments, the study identifies that reviewers of test code predominantly focus on testing-specific concerns, such as test coverage, untested execution paths, assertion quality, mocking practices, and test maintainability. Notably, defects discussed in test reviews tend to be higher-level and more severe than those typically reported in prior literature on production code reviews~\citep{bacchelli2013expectations, beller2014modern}.
The interview results further reveal that reviewing tests poses distinct challenges compared to reviewing production code. Developers report difficulties due to a lack of contextual information, poor navigation between test and production files, and limited tool support. As a result, reviewers often resort to checking out the code locally to understand test behavior and execution context. The study also highlights organizational and cultural factors, including time pressure and managerial incentives, that discourage thorough test code reviews.


Despite the valuable insights provided by Spadini et al., their findings are rooted in a development context that differs substantially from today’s dominant workflows. The original study was conducted on Gerrit, a platform characterized by a patch-based, pre-commit review model with mandatory reviewer involvement and limited automation support~\citep{milanesio2013learning}. In contrast, contemporary software development is largely centered around GitHub, where PRs enable more flexible, negotiation-driven review processes and where reviewer participation is often optional~\citep{rahman2014insight, liu2016comparative}. Moreover, the widespread adoption of GHA has fundamentally altered the review landscape by automating many checks that reviewers previously performed manually, such as test execution, style enforcement, and static analysis~\citep{wessel2023github, decan2022use}. These changes raise important questions about whether and how developers continue to review test code when automated CI pipelines provide immediate feedback. As a result, it remains unclear to what extent the observations reported in the \textit{original study}~\citep{spadini2018testing} generalize to modern platforms and workflows. This motivates the need for a new empirical investigation that re-examines test code review practices in the context of GitHub pull requests and assesses how CI automation, particularly GHA, reshapes reviewer attention and behavior.

\section{Replication Scope and Methodology}\label{sec3}

Table~\ref{tab:prior_work_limitations} compares the limitations of the \textit{original study}~\citep{spadini2018testing} and how our replication experiment addresses them. 
For example, under \textit{Generalizability}, 
one limitation in the \textit{original study} is that it sampled only three projects. This is addressed in our study by expanding to a larger sample and more programming languages. 
Another limitation concerns \textit{Reproducibility}: the \textit{original study} did not specify how unmerged files and reviews were handled. Our replication addresses this by including files appearing in intermediate but unmerged reviews.
Furthermore, considering the inherent differences between the two platforms mentioned in the \textit{Impact of GHA Adoption} criteria, we extend the \textit{original study} by
focusing on how the integration of GHA influences reviewer practices when reviewing test code. 

\begin{table*}[!tp]
\footnotesize
\centering
\caption{Limitations of the \textit{original study} and how our replication addresses them.}
\label{tab:prior_work_limitations}
\begin{tabularx}{\textwidth}{l >{\RaggedRight}X >{\RaggedRight}X}
\toprule
\textbf{Criteria} & \textbf{Limitations in the Original Work} & \textbf{Addressed in This Study} \\
\midrule

\textbf{Generalizability} & 
\begin{itemize}[leftmargin=*, nosep, before=\vspace{0pt}, after=\vspace{0pt}]
  \item Gerrit is adopted mostly by large organizations, focusing on infrastructure and enterprise systems.
  \item Prior work sampled only three projects: Eclipse (Java), OpenStack (Python), and Qt (C++).
  \item Original study lacks project-level analyses with individual examinations.
\end{itemize} & 
\begin{itemize}[leftmargin=*, nosep, before=\vspace{0pt}, after=\vspace{0pt}]
  \item GitHub is the leading, widely used platform (100M+ developers).
  \item Our sample includes: Pandas, Moby, Spark, Flink, VSCode, and TensorFlow (six different languages).
  \item Our project-level analysis uncovers critical heterogeneities and longitudinal shifts that remain hidden in aggregate platform-wide data due to statistical noise.
\end{itemize} \\ 
\addlinespace[8pt] 

\textbf{Reproducibility} & 
\begin{itemize}[leftmargin=*, nosep, before=\vspace{0pt}, after=\vspace{0pt}]
  \item Unknown specific keywords used by authors and bots for filtering cases.
  \item Unclear handling of files modified in intermediate commits but absent from final records.
  \item Lack of standard definitions or a codebook for discussion categories.
\end{itemize} & 
\begin{itemize}[leftmargin=*, nosep, before=\vspace{0pt}, after=\vspace{0pt}]
  \item We developed a keyword list for bots and files inspired by~\citep{golzadeh2020bot}.
  \item Our study explicitly processed intermediate commits to retain files and comments.
  \item We provided explicit descriptions and a codebook for test code review categories.
\end{itemize} \\
\addlinespace[8pt]

\textbf{Impact of GHA} & 
\begin{itemize}[leftmargin=*, nosep, before=\vspace{0pt}, after=\vspace{0pt}]
  \item CI adoption timelines were not considered in previous analysis.
  \item Potential effects of CI on reviewer behavior remained unanalyzed.
\end{itemize} & 
\begin{itemize}[leftmargin=*, nosep, before=\vspace{0pt}, after=\vspace{0pt}]
  \item We explicitly analyzed the impact of GHA adoption on test-related reviews.
  \item We employ both quantitative and qualitative approaches for analysis.
\end{itemize} \\

\bottomrule
\end{tabularx}
\end{table*}

\subsection{Research Questions}\label{subsec3.1}
Our replication was guided by three research questions (RQs). Each RQ focuses on replicating the \textit{original study} on review practices for test code, and on extending this line of inquiry by examining whether adopting GHA alters the observed behaviors.
\\

\noindent\textbf{RQ1: How is reviewers’ attention distributed between production and test code, and does GHA alter this distribution?} 
We investigate whether GitHub displays the same reviewer tendency toward test and production files reported in prior Gerrit-based research. Specifically, we employ a regression analysis to investigate whether there is a statistically observable shift in review patterns for test and production files, comparing datasets both before and after GHA adoption.
\\

\noindent\textbf{RQ2: What topics characterize test code review discussions, and does GHA adoption alter these topics?} 
We first developed a detailed codebook in our replication package~\citep{gha_test_review_2026} to systematically categorize the topics discussed by reviewers concerning test code. This codebook contains definitions and examples of each topic category by the first and second authors after 6 rounds of pair coding to ensure consistency and reliability in our qualitative analysis. To assess the impact of GHA, we applied this codebook to categorize comments from both pre- and post-GHA periods, subsequently comparing the topic distributions to identify statistically significant changes.. 
\\

\noindent\textbf{RQ3: Do reviewers examine test code before production code during code review, and does GHA adoption alter this order?} 
In the \textit{original study}~\citep{spadini2018testing}, the authors interviewed 12 participants and found that reviewers tend to either review test code first or production code first, with different expectations. In our study, we are interested in the distribution of this preference, specifically analyzing from a large-scale statistical perspective whether reviewers are more likely to comment on test files or production files first, and to what extent this tendency persists. We measured this by identifying whether the first comment from a reviewer in a PR targets a test file or a production file. This approach mitigates the bias introduced by interviews in the \textit{original study}. Moreover, we examine whether adopting GHA influences the first-comment distribution, assess whether GHA affects reviewers’ code review practices, and determine whether the review preference is related to code churn.

\subsection{Data Processing}\label{subsec3.2}

This section describes how we constructed and prepared the dataset used in our replication study. We detail the selection of projects, the extraction of pull request data, and the cleaning and filtering steps applied to ensure consistency with the original study and suitability for analyzing test code review practices on GitHub.

\subsubsection{\textbf{Project Selection}}

To investigate current review practices concerning test code in GitHub PRs, we followed the approach of the \textit{original study}~\citep{spadini2018testing} and selected projects according to four criteria: (1)~the repositories have a well-established testing infrastructure; (2) they have a substantial number of PRs, reflecting an established review culture; (3) they remain active and influential in the community; and (4) they have a long-term record of GHA usage. 
To meet these criteria, we first selected the top 495 candidate projects based on forking activity as of Jan 2025, sorting them in descending order. Forking activity, particularly when forks include commit records, often indicates that developers are more likely to engage with the project and submit PRs~\citep{cosentino2017systematic, zhou2014will}. 
Next, using the GitHub RESTful API~\citep{github_api}, we collected the number of PRs for each project and ranked them in descending order. 
From this set, we selected projects that satisfied the additional criteria of having well-established testing infrastructure, demonstrating a certain level of community influence, and maintaining at least 375 days of activity without GHA adoption, as well as 375 days with GHA adoption. 

Ultimately, we selected six projects, each with a different predominant programming language:\footnote{\href{https://innovationgraph.github.com/global-metrics/programming-languages}{https://innovationgraph.github.com/global-metrics/programming-languages}} Pandas (Python), Moby (Go), Spark (Scala), Flink (Java), VSCode (TypeScript), and TensorFlow (C++). We combined data from six projects into a unified dataset as the basis of our analysis. Table~\ref{tab:characteristics_of_Projects} lists the characteristics of the six projects.
Altogether, these projects have 213,511 PRs that were used in our analysis. 

\begin{table*}[t]
\centering
\footnotesize
\caption{Characteristics of Projects}
\label{tab:characteristics_of_Projects}

\newcolumntype{R}{>{\raggedleft\arraybackslash}X}

\setlength{\tabcolsep}{3pt}      
\renewcommand{\arraystretch}{0.95} 


\makebox[\textwidth][r]{%
\begin{tabularx}{\textwidth}{l*{7}{R}}
\toprule
Project & Pandas & Moby & Spark & Flink & VSCode & Tensorflow & Total \\
\midrule
Primary Language & Python & Go & Scala & Java & TypeScript & C++ & - \\
\rowcolor[gray]{0.95}
\# of PR & 33,535 & 25,886 & 49,489 & 25,983 & 35,679 & 42,929 & 213,511 \\
\# of Comments & 114,421 & 57,798 & 253,798 & 110,686 & 24,796 & 39,406 & 600,905 \\
\rowcolor[gray]{0.95}
\# of PRs Retained & 14,440 & 8,157 & 25,201 & 11,147 & 4,628 & 5,319 & 68,892 \\
Cum. Test Code Churn & 1,828,164 & 813,356 & 3,280,282 & 3,289,469 & 499,423 & 4,731,039 & 14,441,733 \\
\rowcolor[gray]{0.95}
Cum. Prod. Code Churn & 3,682,770 & 6,912,697 & 8,128,655 & 5,132,898 & 3,691,608 & 8,879,857 & 36,428,485 \\
\# of contributors & 3,427 & 2,271 & 2,167 & 1,303 & 3,688 & 2,166 & 15,022 \\
\rowcolor[gray]{0.95}
GHA Adoption Time & 11/16/19 & 03/14/22 & 08/13/19 & 08/23/23 & 10/21/19 & 10/05/20 & - \\
\bottomrule
\end{tabularx}%
}
\end{table*}

\subsubsection{\textbf{Data Cleaning}}

To ensure quality and maintain consistency with the \textit{original study}~\citep{spadini2018testing}, we applied the following exclusion criteria when analyzing \textbf{RQ1}: (1) \textit{Interaction-based Filtering}, PRs were excluded if they contained no comments, or if the discussion was limited exclusively to automated bots and the PR author; (2) \textit{Content-based Filtering}, PRs consisting solely of changes to documentation or configuration files were removed; and (3) \textit{Comment Handling}, comments associated with outdated file\footnote{Example of Outdated File: \href{https://github.com/pandas-dev/pandas/pull/46860}{https://github.com/pandas-dev/pandas/pull/46860}} versions were treated as standard comments to preserve the full context of the code review process.
The final dataset comprised 68,892 PRs (\# of PRs Retained in Table~\ref{tab:characteristics_of_Projects}), each containing at least one reviewer comment and involving modifications to at least one production or test file, forming the basis for addressing \textbf{RQ1}.

To investigate the impact of GHA adoption on test code review, we treat CI configuration files as the point of CI adoption, following the approach proposed by ~\citep{golzadeh2022rise}. 
Consequently, we determined the dates when the GHA configuration was first introduced in each of the six selected projects and reported these results as \textit{GHA Adoption Time }in Table~\ref{tab:characteristics_of_Projects}. Comments happened in test files were extracted and sampled to answer \textbf{RQ2}.

To evaluate reviewers' initial focus, we applied the following inclusion criteria for \textbf{RQ3}: (1) PRs had to modify at least one test file and at least one production file; (2) PRs had to include at least one review comment on either a test file or a production file; and (3) PRs had to fall within a 24-month window before and after GHA adoption. The resulting dataset comprised 6,166 PRs for \textbf{RQ3}.

\begin{figure}[htbp]
    \centering
    \includegraphics[width=0.9\textwidth]{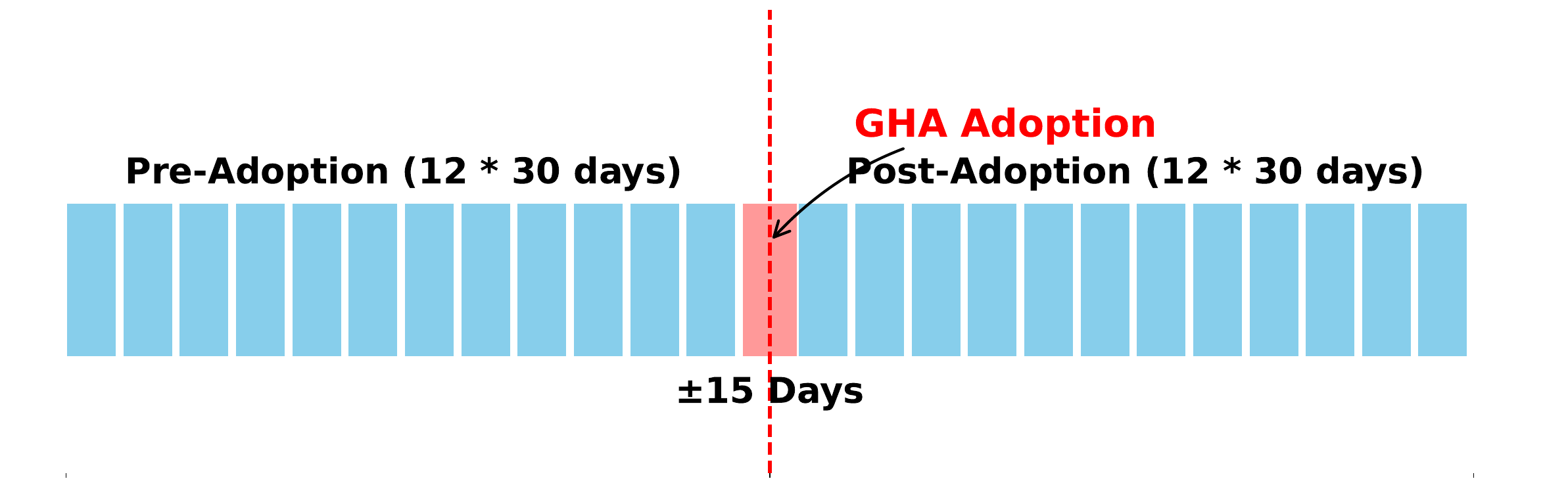}
    \caption{Overview of the Time-based RDD Methodology.}
    \label{fig:ITSA_Methodology}
\end{figure}


\subsection{Evaluation Metrics}\label{subsec3.3}
Following the methodology of the \textit{original study}~\citep{spadini2018testing}, we analyzed GitHub PRs that contain at least one reviewer comment and modify either test files, production files, or both. Additionally, we included outdated files in our analysis. Specifically, we identify outdated files as files modified during review but excluded from the final merge. Review comments linked to these files are incorporated into our analysis. A single file may be counted multiple times if it was modified in different PRs across the project history.

\subsubsection{RQ1 Considerations}
In \textbf{RQ1}, we quantify reviewers’ relative attention to test code using the odds ratio (OR), computed as the odds that production files receive comments divided by the odds that test files receive comments. Values greater than 1 indicate that reviewers are more likely to comment on production code than on test code. To align with the \textit{original study}, we follow the same data processing pipeline, which facilitates a more reliable assessment of whether the reported patterns generalize across platforms. Specifically, we calculate ORs under two conditions: (1) PRs that include changes to both production and test files; and (2) PRs that include changes exclusively to production files or exclusively to test files.

We further investigate whether adopting GHA is associated with changes in review practices over time~\citep{golzadeh2022rise}. Compared to the pre-adoption period, GHA can automate and standardize test execution and related checks within the pull request workflow, potentially altering how reviewers allocate their attention during code review. This motivates the question of whether reviewers’ focus shifts after GHA adoption (e.g., away from test-execution-related concerns and toward higher-level issues).
To examine this question, we employ a time-based regression discontinuity design (RDD)~\citep{trochim1990regression, hahn2001identification, zhao2017impact}, a quasi-experimental method in which a specific cutoff determines who receives an intervention. We use each project’s GHA adoption date as a cutoff date and align the projects by event time. Our model accounts for project-level time-invariant heterogeneity, overall temporal trends, post-adoption slope changes, file composition (test vs. production changes), and monthly pull request volume. To reduce potential instability during the transition, we exclude the 15-day periods immediately before and after adoption, as shown in Figure~\ref{fig:ITSA_Methodology}. For each project, we collect 12 consecutive 30-day intervals before and after adoption, and then aggregate the aligned intervals across projects to enable cross-project analysis of both short- and long-term effects around the adoption cutoff.

\begin{equation}
\label{eq:gha_impact}
\begin{aligned}
\ln(OR) \sim \, & \beta_0 + \beta_1 \textit{Time} + \beta_2 \textit{postGHA} + \beta_3 \textit{TimeAfter} \\
& + \beta_4 \textit{mean\_test\_mod\_ratio} + \gamma C(\textit{Project}) + \epsilon
\end{aligned}
\end{equation}

To evaluate the impact of GHA adoption on review preferences, we formalize our analysis using the regression model presented in Equation~\eqref{eq:gha_impact}. The dependent variable, $ln(\text{OR})$, represents the log odds ratio of the review preference. The model incorporates the following components:

\textbf{Interrupted Time Series Parameters:} $\beta_1$ ($\mathit{Time}$) captures the baseline monthly trend prior to the intervention; $\beta_2$ ($\mathit{postGHA}$) estimates the immediate level shift (step change) following GHA adoption; and $\beta_3$ ($\mathit{TimeAfter}$) reflects the change in the trend slope in the post-adoption period.

\textbf{Key Covariate:} $\beta_4$ ($\mathit{mean\_test\_mod\_ratio}$) serves as a proxy for testing intensity, calculated as the monthly average proportion of test files modified per PR.

\textbf{Project Controls:} $C(\mathit{Project})$ denotes project-specific fixed effects. This term accounts for time-invariant heterogeneity across the studied repositories, effectively controlling for factors such as primary programming language, team size, and project maturity that remain relatively stable throughout the observation period.

A critical step in our model construction involved addressing potential conceptual and mathematical overlap between the confounding variables and the dependent variable. We conducted a redundancy analysis following the guidelines by~\citep{thongtanunam2017review} and observed that certain metrics, such as monthly average proportion of test files been comments per PR, were highly correlated with the dependent variable underlying logic. Including such parameters would lead to spurious correlation and an artificially inflated $R^2$, as the model would be explaining the variance through redundant mathematical identities rather than actual behavioral shifts. To ensure the model's interpretability and prevent overfitting, we excluded overlapping parameters, retaining only $\mathit{mean\_test\_mod\_ratio}$ to maintain parsimony and robustness. Code churn related parameters were excluded to avoid multicollinearity, as its informational value significantly overlaps with the existing metrics.


The results of the experiments above demonstrate that project-level mixed effects are a key factor in shaping review practices (further details in Section~\ref{sec4}). To better capture the long-term impact of GHA on reviewer practices, we examined whether two key indicators changed over time before and after GHA adoption for each project on a monthly basis: (1) the \textit{``test file review rate"}, defined as the ratio of modified test files that received at least one comment to the total number of modified test files within a PR; and (2) the \textit{``test file review density"}, defined as the total number of comments on test files divided by the number of modified test files within a PR. 
As preliminary experimental results indicate that both indicators are heavily skewed~\citep{kalliamvakou2014promises}, we evaluated the results using the medians of the two indicators, calculated within 30-day intervals.

\subsubsection{RQ2 Considerations}
For \textbf{RQ2}, we extracted all review comments from our dataset within the 360 days before and after the adoption of GHA (unstable period excluded). From this pool, we randomly sampled 385 comments from the pre-GHA period and 385 comments from the post-GHA period and annotated them separately. This sample size was determined according to the guidelines of Krejcie and Morgan~\citep{krejcie1970determining}, ensuring a confidence level of 95\% with a margin of error of 5\% for the adoption of GHA before and after separately, totally achieving a confidence level of 99\% with a margin of error of 5\% for the combined dataset on GitHub.

To calibrate the annotation framework prior to the full-scale analysis, the first and second authors conducted six iterative coding rounds on a fixed pilot set of 70 review comments (35 pre- and 35 post-GHA adoption). In each round, both authors independently applied the provisional category scheme and subsequently reconciled our annotations through negotiated agreement~\citep{wicks2017coding}. The initial round yielded an inter-rater agreement of only 50.7\% among 18 sub-categories, indicating considerable ambiguity in category interpretation. Agreement reached 77.0\% by the sixth round, at which point the codebook was finalized and consensus was established. Based on this validated coding scheme, the remaining 350 review comments (175 pre- and 175 post-GHA adoption) were evenly divided between the two authors for independent annotation, with residual uncertainties from the jointly annotated subset revisited to ensure consistency and reproducibility across the entire dataset~\citep{gha_test_review_2026}.

To examine whether the distribution of comment categories changed before and after the adoption of GHA, we conducted a Chi-square test of independence~\citep{pearson1900x}. 

\subsubsection{RQ3 Considerations}

Whereas the \textit{original study}~\citep{spadini2018testing} relied on interviews to identify two common practices, we employed a quantitative approach to investigate these practices on GitHub. Specifically, we analyzed the timestamps of review comments within individual PRs to determine whether the first comment was made on a test file or a production file, which we interpret as indicative of the reviewer’s initial focus. 
We further examined whether this order of review is associated with code churn (e.g., whether reviewers are more likely to review test files first when test code churn exceeds production code churn in the same PR) and whether adopting GHA influenced these tendencies. This large-scale comment analysis, spanning six high-impact projects and hundreds of thousands of PRs, offers scalability and reproducibility beyond what interview-based methods can provide, though future research may integrate both approaches to yield a more holistic understanding of reviewer practices.

\begin{equation}
\label{eq:test_review_impact}
\begin{aligned}
\text{test\_review\_first\_ratio} \sim \, & \beta_0 + \beta_1 \textit{PostGHA} + \beta_2 \textit{Time} + \beta_3 \textit{TimeAfter} \\
& + \beta_4 \textit{mean\_pr\_files} + \beta_5 \textit{mean\_test\_churn} + \beta_6 \textit{mean\_prod\_churn} \\
& + \gamma C(\textit{Project}) + \epsilon
\end{aligned}
\end{equation}

To evaluate the impact of GHA, we constructed a regression model (Equation~\ref{eq:test_review_impact}) that accounts for confounding variables, including \textit{mean\_prod\_churn}, \textit{mean\_test\_churn}, and the number of files per PR calculated as monthly averages. To control for inherent differences between subjects, we also incorporated project fixed-effects. We first extracted a dataset of 64,662 PRs spanning a 48-month window surrounding GHA adoption. From these, we further selected PRs that modified both production and test files and that received at least one comment on either test files or production files, yielding 6,166 PRs for examining reviewers’ preferences in the order of review. Review order is an important consideration as prior research demonstrates that the relative position of files in the code review interface can influence developers’ attention allocation and defect detection, thereby affecting the overall outcome of code reviews~\citep{fregnan2022first, desai2025priotestci}.



\begin{table}[t]
\centering
\footnotesize 
\caption{Prevalence of Reviews in Test vs. Production Files}
\label{tab:review_prevalence}
\setlength{\tabcolsep}{5pt}
\begin{tabular}{@{} l r r r c r r @{}}
\toprule
\textbf{File} & 
\textbf{\# Files} & 
\textbf{w/ Comm.} & 
\textbf{w/o Comm.} & 
\textbf{Odds Ratio} & 
\textbf{\# Comm.} & 
\textbf{Avg. Comm.} \\ 
\textbf{Category} & 
\textbf{} & 
 & 
 & 
\textbf{(Gerrit)} & 
\textbf{} & 
\textbf{(Gerrit)} \\ 
\midrule

\multicolumn{7}{l}{\textit{Scenario A: PRs containing both Production and Test files}} \\
\midrule
Production & 185,660 & 66,657 (35.90\%) & 119,033 (64.10\%) & \multirow{2}{*}{\textbf{1.54} (1.90)} & 284,998 & 1.54 (3.00) \\
Test       & 104,898 & 28,022 (26.71\%) & 76,876 (73.29\%)  &  & 90,690 & 0.86 (1.27) \\

\midrule
\addlinespace[1.5ex] 

\multicolumn{7}{l}{\textit{Scenario B: PRs containing exclusively Production or Test files}} \\
\midrule
Only Prod. & 73,098  & 28,703 (39.27\%) & 44,395 (60.73\%)  & \multirow{2}{*}{\textbf{1.05} (0.86)} & 109,184 & 1.49 (1.64) \\ 
Only Test  & 18,390  & 7,000 (38.06\%)  & 11,390 (61.93\%)  &  & 26,174 & 1.42 (2.30) \\
\bottomrule
\addlinespace[1ex]
\multicolumn{7}{l}{\textit{Note:} 1.90 and 0.86 under Odds Ratio column represent reference Gerrit data.}
\end{tabular}
\end{table}

\section{Result}\label{sec4}
In the following sections, we present the results of our replication study, along with insights and implications. The results and analysis are organized per RQs.

\subsection{RQ1: How is reviewers’ attention distributed between production and test code, and does GHA alter this distribution?}

For this RQ, we analyze how reviewers’ attention in terms of discussion distribution differs between Gerrit and GitHub, and how it changes before and after the adoption of GHA.

\subsubsection{\textbf{Platform Level Comparison: Gerrit vs. GitHub}}

Table~\ref{tab:review_prevalence} presents the descriptive statistics for the dataset constructed in Section~\ref{sec3}. Following the methodology of the \textit{original study}, we categorized the results into two scenarios. In Scenario A, which covers all PRs containing both file types (i.e., production and test), we observed 185,660 production and 104,898 test file modifications. Specifically, 66,657 production files and 28,022 test files received 284,998 and 90,690 comments, respectively. This yields an odds ratio of 1.54, compared to 1.90 in the original Gerrit-based study. Furthermore, the average number of comments per production file, in the last column of the table, computed as $(\# of Comm.) / (\# Files)$, was 1.54 (vs. 3.00 in Gerrit), while test files received an average of 0.86 comments (compared to 1.27 in Gerrit). Similarly, the statistics for Scenario B are interpreted using the same metrics, on GitHub, the OR is 1.05 which is different in Gerrit (OR = 0.86), in which test code has a higher probability of being reviewed than production code in Gerrit.
Together, these results suggest that GitHub PRs exhibit a more balanced distribution of review activity across test and production code. 
We found several factors likely to contribute to this shift, including the growing awareness of developers of the value of test code~\citep{spadini2019test}, community norms that explicitly encourage its review~\citep{zhang2022consistent}, and the GitHub user interface that provides an easier way to navigate through the test and tested classes~\citep{spadini2018testing}. 

\begin{figure*}[t] 
    \centering
    
    \begin{subfigure}[t]{0.90\textwidth} 
        \centering
        \includegraphics[width=\linewidth]{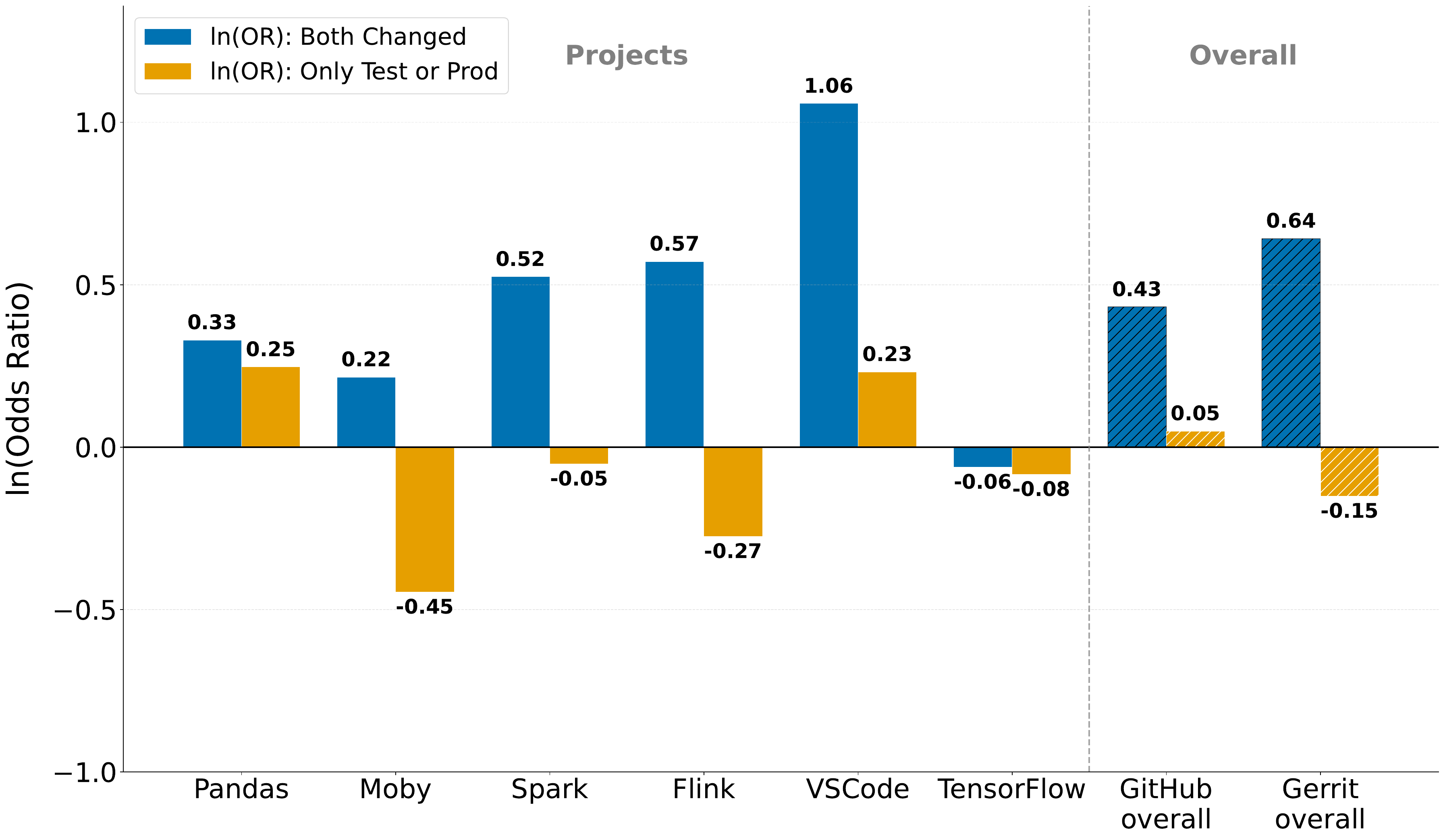}
        \caption{Project-level comparison}
        \label{fig:github_vs_gerrit_32pt}
    \end{subfigure}

    \begin{subfigure}[t]{0.90\textwidth}
        \centering
        \includegraphics[width=\linewidth]{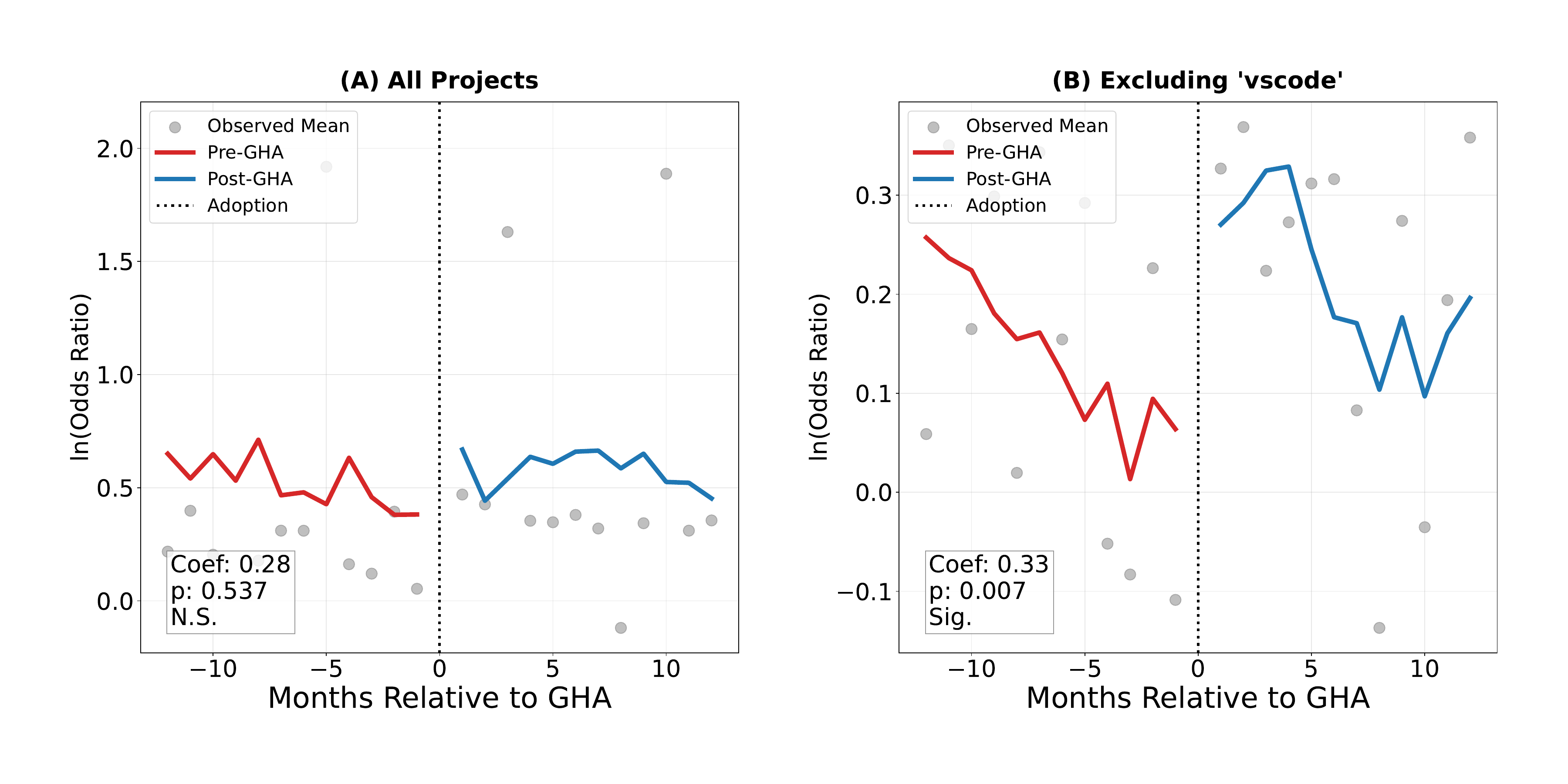}
        \caption{Relationship between Odds Ratio and GHA adoption.}
        \label{fig:rdd_gha_odds_ratio_plot}
    \end{subfigure}

    \caption{Analysis of code review prevalence and GHA impact. (a) Highlights the variance across projects, while (b) illustrates the correlation with GHA adoption.}
    \label{fig:combined_analysis}
\end{figure*}

Figure~\ref{fig:github_vs_gerrit_32pt} presents the OR for both project level and platform level, where the baseline $ln(\text{OR}) = 0$ corresponds to no bias. Project-level results are shown as solid bars on the left, while aggregate platform-level results (GitHub and Gerrit overall) are shown using diagonally hatched bars on the right. 
At the platform level, we observe that review practices on GitHub PRs are overall more balanced (meaning that the review effort is more evenly distributed between test and production code compared to Gerrit, $ln(\text{OR})$ closer to zero). At the project level, the blue bars represent the $ln(\text{OR})$ for Scenario A. 
The yellow bars, by contrast, reflect the results for Scenario B. 
Our analysis reveals substantial heterogeneity in review practices across projects. For example, in Scenario A, the $ln(\textit{OR})$ for TensorFlow ($ln(\textit{OR})$ = -0.06) and VSCode ($ln(\textit{OR})$ = 1.06) point in opposite directions, while Moby ($ln(\textit{OR})$ = -0.45) shows a distinct preference to review test files in Scenario B. 
Consequently, treating all projects as a monolithic entity yields only a broad statistical average, risking the obscuration of these critical nuances. 

We also note that the average number of comments per file (\textit{Avg \# of comment} in Table~\ref{tab:review_prevalence}) is consistently lower on GitHub than on Gerrit. This difference is not surprising; rather, it reflects the distinct motivations underlying the review models used by the two platforms. Gerrit enforces code review before changes are merged into the main history, thereby establishing a strong quality gate. By contrast, GitHub’s review model is designed around PRs on feature branches, where reviews typically occur after code has already been committed, and reviewers who deem the code satisfactory often leave no comments. Thus, the observed divergence in outcomes aligns naturally with the contrasting philosophies of the two models. A similar finding is observed in the work of Pandya et al.~\citep{pandya2022corms}. Although their study did not emphasize this result due to differing research objectives, we reprocessed the data reported in their Table 8 and found that projects on the Gerrit platform exhibit a much higher code review probability (the ratio of total reviews to total modified files, $\approx0.1749$) compared to those on the GitHub platform ($\approx0.0471$), confirming the trend reported in our analysis.

\begin{tcolorbox}[colback=black!5,
                  colframe=black,
                  rounded corners=southwest,
                  rounded corners=northwest,
                  boxrule=0.5pt,
                  arc=8pt,
                  left=5pt,
                  right=5pt,
                  top=5pt,
                  bottom=5pt,
                  fonttitle=\bfseries,
                  before skip=10pt,
                  after skip=10pt,
                  breakable]
\textbf{Finding 1.} \textit{Compared to Gerrit, PRs on GitHub are less likely to receive review comments overall, but exhibit a more balanced distribution between test and production files. At the same time, review practices vary considerably across projects, showing that both platform models and project-specific factors shape how test code is reviewed.}
\end{tcolorbox}

\textbf{Actionable implications:}
Based on \textbf{Finding 1}, future code review research should factor in project and platform-level variations rather than assuming uniform reviewer behavior. Secondly, we found that the review probability on GitHub is significantly lower, indicating that project maintainers cannot rely on the platform's default mechanisms to ensure code is thoroughly reviewed. On the contrary, they may need to develop review guidelines or introduce tools to increase the likelihood that code is reviewed. Finally, the heterogeneity across projects indicates that review practices are influenced not only by platform models but also by project-specific factors (i.e., variables), such as programming languages, tacit review conventions that evolve organically among contributors, and project maturity, underscoring the importance of tailoring review processes to project goals and development culture. 

\subsubsection{\textbf{GHA Influence: Odds Ratio Perspective}}\label{sec4.1.2}

Figure~\ref{fig:rdd_gha_odds_ratio_plot} presents the evolution of the $ln(\text{OR})$ around the GHA adoption cutoff. Using the sample with all projects, the regression analysis (Figure~\ref{fig:rdd_gha_odds_ratio_plot} left) shows no statistically significant discontinuity at the point of GHA adoption (coef = 0.283, p~=~0.537). This result suggests that, on average, GHA adoption is not associated with a common or systematic shift in reviewers’ relative attention between production and test code across projects. 

The lack of a statistically significant effect does not imply uniformity across the sample. During the analysis, we observed substantial project-level heterogeneity. In particular, the fixed effect coefficient for VSCode is large and statistically significant ($\beta$ = 3.209, p = 0.010), indicating that its review attention pattern differs markedly from that of other projects. Also, the VSCode project shows extremely low attention to test code, with only 236 out of 24,796 reviewer comments on test files throughout the project’s history. Such a pronounced and persistent deviation suggests that VSCode may follow a distinct review regime, potentially inconsistent with the parallel-trends assumption underlying our identification strategy.
Motivated by this heterogeneity, we re-estimate (Figure~\ref{fig:rdd_gha_odds_ratio_plot} right) the model after excluding VSCode as an outlier. In this specification, we observe a statistically significant positive discontinuity at the GHA adoption cutoff ($\beta$  = 0.329, p = 0.007), indicating an immediate upward shift in $ln(\text{OR})$. In contrast, the post-adoption slope change remains negligible, suggesting that the effect is driven primarily by a short-term level shift rather than a change in long-term trends. 

The observed discontinuity is unsurprising. Prior to GHA adoption, projects often focus on maintaining test code in preparation for the transition, while immediately after adoption, reviewers shift attention to production code, as GHA exposes defects that require targeted adjustments.




\begin{tcolorbox}[colback=black!5,
                  colframe=black,
                  rounded corners=southwest,
                  rounded corners=northwest,
                  boxrule=0.5pt,
                  arc=8pt,
                  left=5pt,
                  right=5pt,
                  top=5pt,
                  bottom=5pt,
                  fonttitle=\bfseries,
                  before skip=10pt,
                  after skip=10pt,
                  breakable]
\textbf{Finding 2.} \textit{Project-level effects captured by the mixed-effects model appear to be the primary determinants of reviewer attention compared with GHA adoption. After excluding VSCode, the adoption of GHA had a statistically significant immediate effect, making reviewers more likely to focus on production code, but no significant long-term trend was observed.}
\end{tcolorbox}

\textbf{Actionable implications:}
For researchers, the results of Finding 2 underscore the need to explicitly model project-level heterogeneity when studying code review behavior: reviewer attention is driven far more by entrenched project practices than by CI automation adoption alone. 
GHA’s transient influence suggests a `novelty effect' that rapidly fades into complacency. As practitioners increasingly leverage LLMs for complex test generation, the risk of `over-reliance' intensifies. If simple CI triggers prompt reviewer fatigue, the high-density output of AI-generated tests~\citep{chen2024chatunitest, schafer2023empirical} combined with CI will likely lead to even greater oversight, as reviewers may lack the cognitive bandwidth or motivation to scrutinize vast quantities of AI-produced code.
Team leads should proactively recalibrate review guidelines after introducing CI automation, ensuring that test code does not become implicitly ``trusted by default'' once checks pass. This may include explicit expectations for test review or separating test-focused review phases~\citep{li2022follow, zhang2022consistent}. For tool builders, the absence of a long-term trend highlights an opportunity: CI systems currently influence reviewer behavior only transiently. Tooling that persistently surfaces test quality signals (e.g., test relevance, coverage gaps, or historical test fragility) at review time could counterbalance production-centric attention and better align reviewer effort with long-term quality risks.

\subsubsection{\textbf{GHA Influence: Review Rate \& Review Density Perspective}}

\begin{figure}[htbp] %
    \centering

    \begin{subfigure}{\linewidth}
        \centering
        \includegraphics[width=0.95\linewidth]{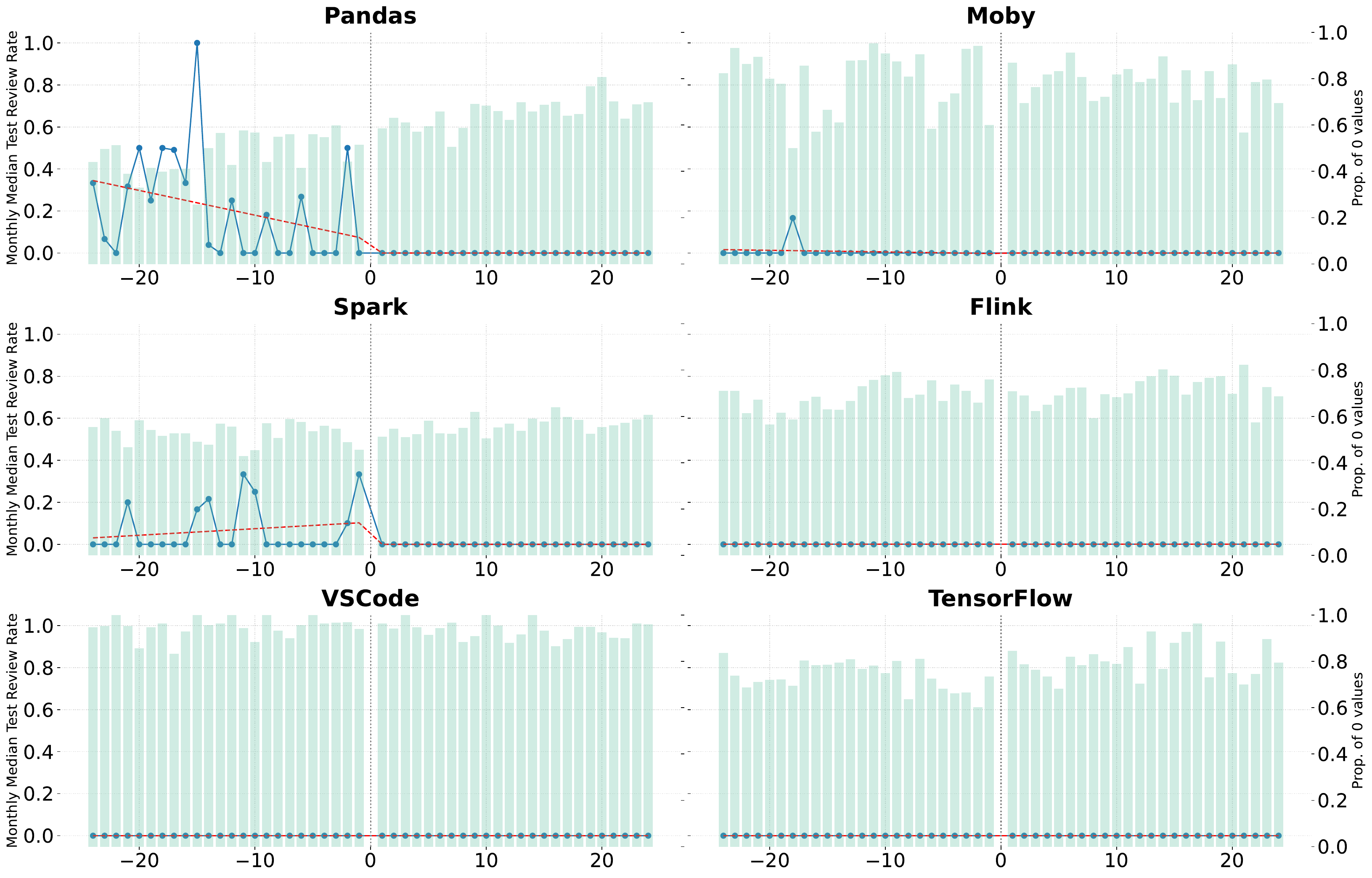}
        \caption{Monthly Medians of Test Files Review Rate.}
        \label{fig:test_review}
    \end{subfigure}

    \begin{subfigure}{\linewidth}
        \centering
        \includegraphics[width=0.95\linewidth]{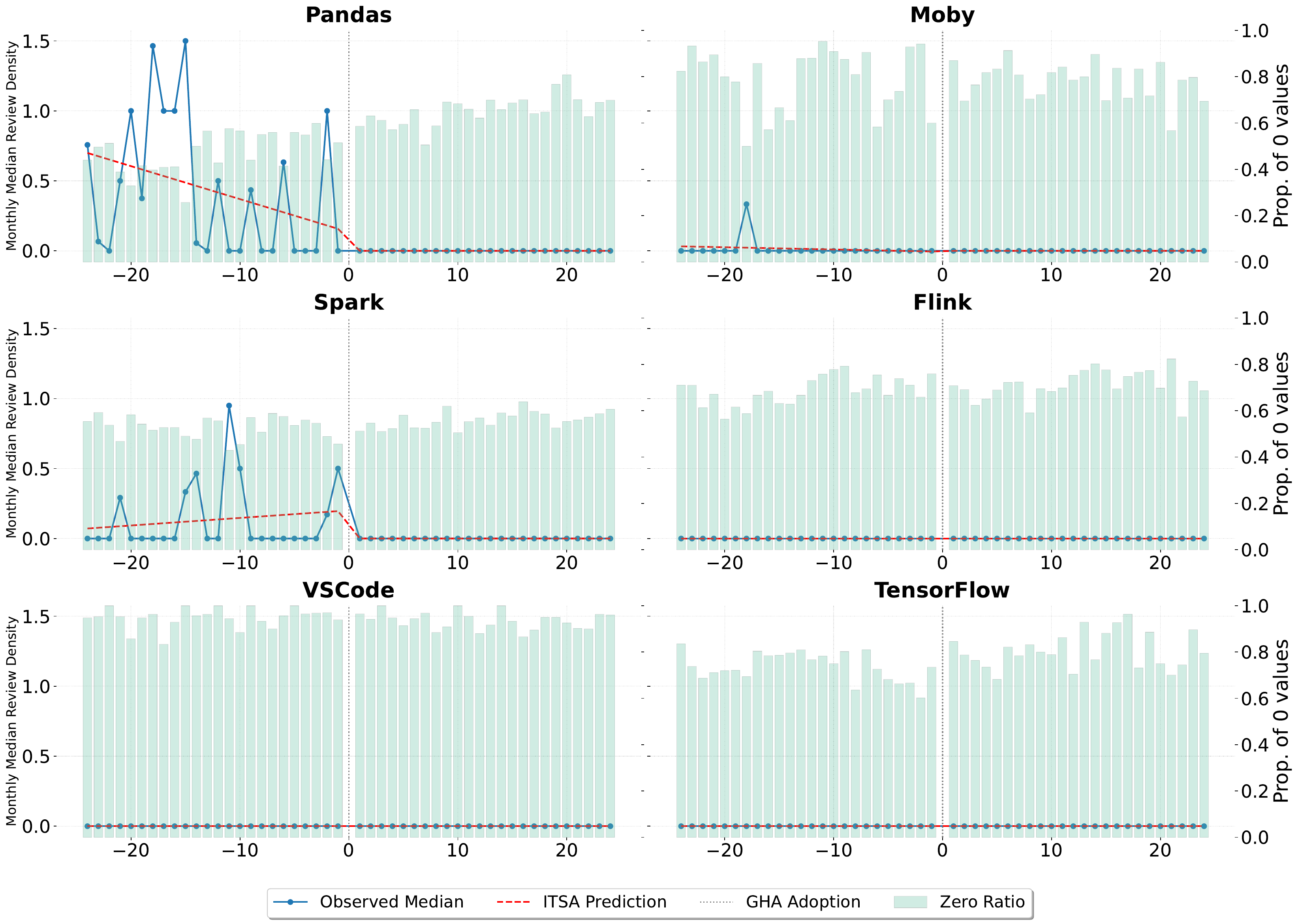}
        \caption{Monthly Medians of Test Files Review Density.}
        \label{fig:itsa_density}
    \end{subfigure}

    \caption{Interrupted Time Series Analysis (ITSA) of test file review metrics before and after GHA adoption. (a) shows the trend in test code review rates, while (b) illustrates the change in test code review density.}
    \label{fig:itsa_dual_comparison}
\end{figure}
Based on the results above, we find that project-level differences play an important role in shaping review practices. Therefore, we conduct separate analyses of test code review activity for each project. Figures \ref{fig:test_review} and~\ref{fig:itsa_density} present the monthly medians of the test file review rate and the test file review density (left Y-axis). As illustrated in the figures, both metrics (represented by blue data points) exhibited a discernible decline following the adoption of GHA in the Pandas and Spark projects. Statistical analysis confirms this observation, revealing a significant immediate level shift characterized by a reduction in test-related code reviews ($p<0.001$ for the intercept). Furthermore, the Spark project demonstrated a sustained longitudinal decline, with the time\_after\_intervention variable indicating a significant long-term downward trend ($p<0.012$). In contrast, for the Flink, VSCode, and TensorFlow projects, the review rate and review density of test code were so low before and after GHA adoption that the median values remained at or near zero, providing little information about potential changes. This outcome is expected given the highly skewed nature of the data: a large proportion of modified test files received no comments, leaving many PRs with review rate and density values of zero. While the skewed distribution justifies the use of medians, in this context, the median is limited in its ability to reflect underlying changes in reviewer behavior. This aligns with our findings in Section~\ref{sec4.1.2}, which explain the absence of a long-term impact following the adoption of GHA. These trends suggest that after the adoption of GHA, in projects with established review cultures like Spark and Pandas, GHA adoption appears to have triggered a substitution effect, displacing manual oversight with automation. Conversely, in projects where test reviews were already infrequent, the intervention further solidified the marginalization of test code scrutiny. Collectively, these findings imply that while CI/CD automation streamlines the development pipeline, it may inadvertently diminish the rigor of manual peer reviews for test-related contributions.

In each chart of Figure~\ref{fig:itsa_dual_comparison}, the green histograms show, for each month, the proportion of PRs that changed test files but received no reviewer comments on the test code (right Y-axis). 
The visualization shows that, across the Pandas, Spark, and TensorFlow projects, the proportion of PRs that changed test files but received no reviews increased after adopting GHA. This pattern provides additional evidence that GHA is associated with reduced reviewer attention to test code, both in terms of frequency (review rate) and intensity (review density).

The VSCode project deserves a separate discussion due to its notably low rate of test code reviews. 
In our replication package, we also analyzed Grafana, another TypeScript-based project, and observed a similarly low level of attention to test code ($ln(\text{OR})$ = 1.02). These findings suggest that the programming language may also be a factor influencing whether projects prioritize test code review, which has already been captured in confound \textit{C(Projects)} in Equation~\ref{eq:gha_impact}. This is a direction warranting further investigation in future work.

\begin{tcolorbox}[colback=black!5,
                  colframe=black,
                  rounded corners=southwest,
                  rounded corners=northwest,
                  boxrule=0.5pt,
                  arc=8pt,
                  left=5pt,
                  right=5pt,
                  top=5pt,
                  bottom=5pt,
                  fonttitle=\bfseries,
                  before skip=10pt,
                  after skip=10pt,
                  breakable]

\textbf{Finding 3.} \textit{Across projects, the adoption of GHA is associated with a significant decline in reviewers’ engagement with test code, reflected by decreasing monthly medians of test file review rate and test file review density in Pandas and Spark. In projects where test review activity is already sparse (Flink, VSCode, TensorFlow), median-based indicators remain at or near zero because of zero inflation, limiting their ability to capture behavioral shifts.}
\end{tcolorbox}

\textbf{Actionable implications:} The observed decline in reviewers’ engagement with test code following GHA adoption on Pandas and Spark suggests that automation can inadvertently displace human oversight on some projects rather than complement it. For practitioners, this implies that a passing CI pipeline should not be treated as a proxy for test quality: teams may need to explicitly mandate human review of test changes (e.g., checklist items, PR templates, or required reviewers for test files) to counterbalance the trust in automation. For projects with already sparse test review activity, relying on aggregate or median-based indicators is insufficient; instead, teams and researchers should adopt alternative signals (e.g., tail-sensitive metrics, qualitative audits, or targeted sampling) to surface subtle but systematic neglect. 
For tool designers, these insights inspire the design of review tools that can monitor the influence of CI and LLMs on team review practice dynamics. By detecting shifts in review practices, these tools can provide dynamic analysis that empowers teams to proactively cultivate a more robust, sustainable review culture aligned with their expectations.

\subsection{RQ2: What topics characterize test code review discussions, and does GHA adoption alter these topics?}

Beyond how often test code is discussed, it is equally important to understand the quality of test code review. In this RQ, we characterize the topics that dominate test code review discussions and examine whether GHA adoption shifts the nature of these conversations.

\subsubsection{\textbf{Platfrom Comparison: Gerrit vs. GitHub.}}

\begin{figure}[!tp]
    \centering
    \begin{subfigure}{\linewidth}
        \centering
        \includegraphics[width=0.9\linewidth]{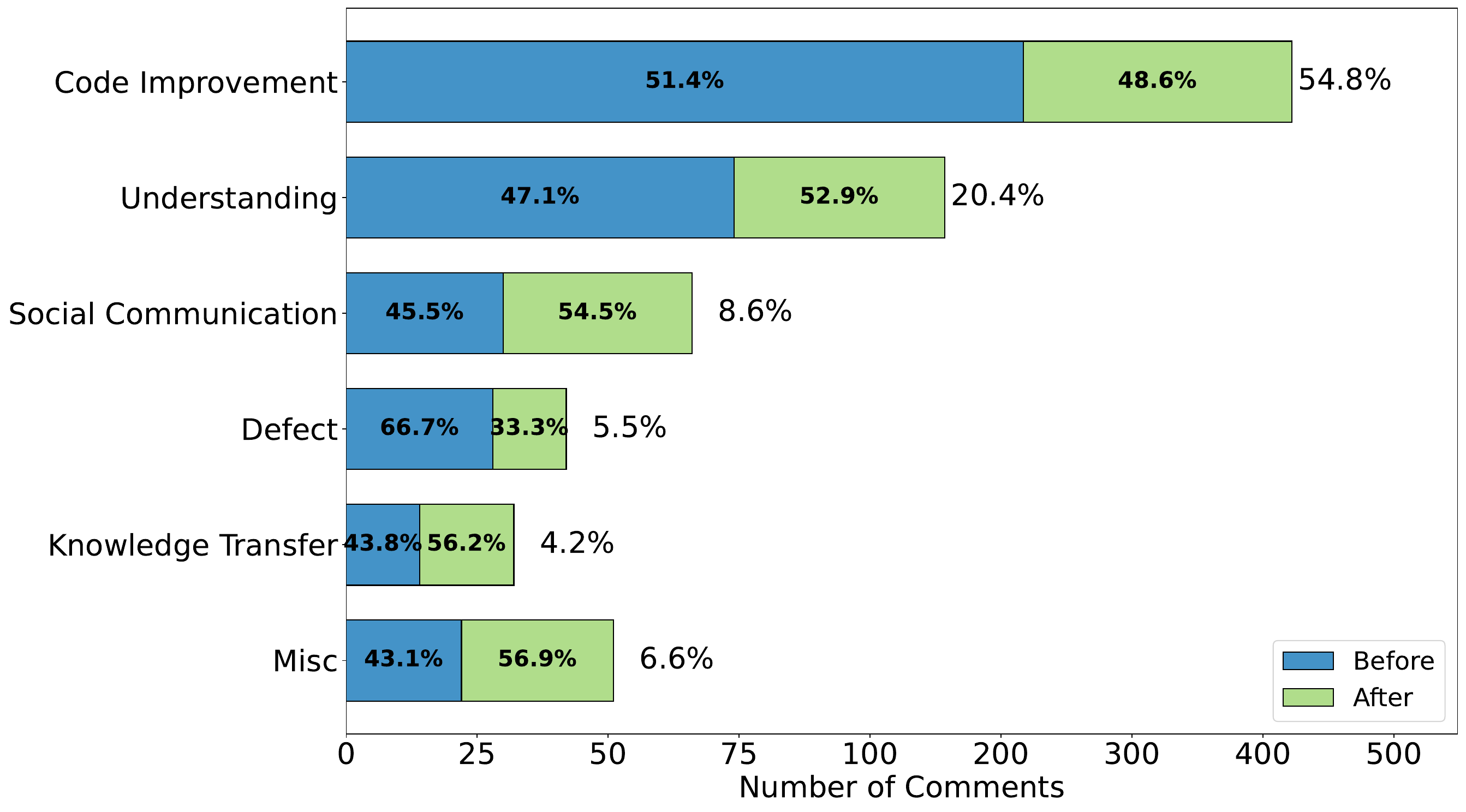}
        \caption{Major Categories Distribution}
        \label{fig:majorCategories}
    \end{subfigure}

    \begin{subfigure}{\linewidth}
        \centering
        \includegraphics[width=0.9\linewidth]{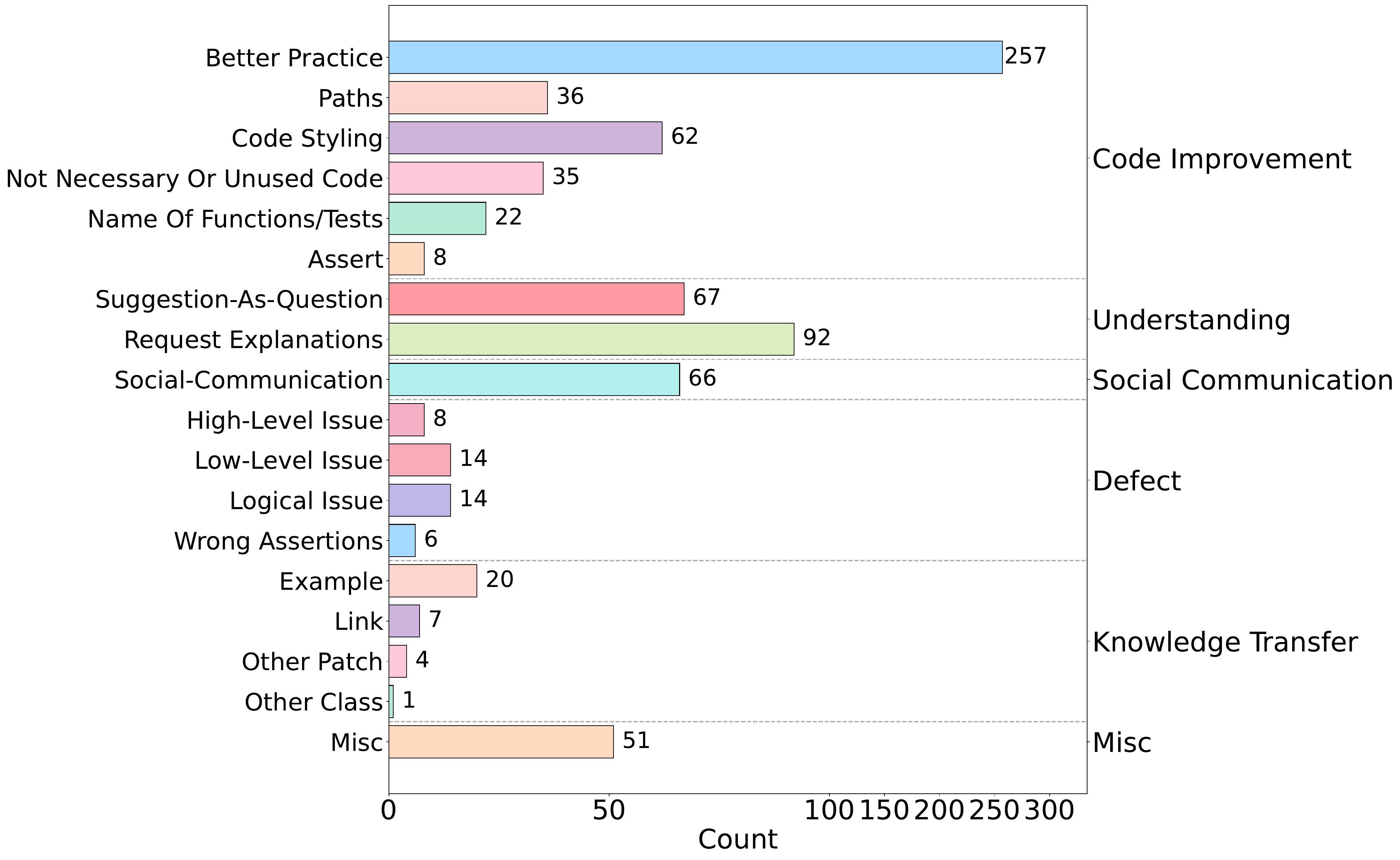}
        \caption{Detailed Subcategory Breakdown}
        \label{fig:subCategories}
    \end{subfigure}
    
    \caption{Distribution and categories of topics identified in the manual analysis of test code reviews ($N=770$). The classification highlights prevalent concerns raised by reviewers during the assessment of test suites.}
    \label{fig:combinedRQ2}
\end{figure}

Figure~\ref{fig:combinedRQ2} presents the six categories of review comments identified in related work~\citep{spadini2018testing, bacchelli2013expectations} (Figure~\ref{fig:majorCategories}) and the subcategories of each major category (Figure~\ref{fig:subCategories}). The former also has a segmented view of the distributions of the categories before and after the adoption of GHA, which together represent the overall distribution of our sampled data. Detailed definitions of these categories, along with the distribution of subcategories, are available in our replication package~\citep{gha_test_review_2026}.

Compared to the findings of the \textit{original study}~\citep{spadini2018testing}, we observed notable structural differences in category distributions across platforms. These discrepancies are attributed to differences in platform mechanisms, review processes, or project-specific practices, as discussed in what follows.

\textbf{Code Improvement (54.8\%):} On the GitHub platform, test-related discussions more frequently focus on code improvement compared to the Gerrit platform (35\%). This increased attention comes at the expense of other categories, such as defect detection and code comprehension. Comments in this category typically encourage authors to enhance performance (e.g., \textit{``Rather than importing this should make this a fixture for reuse in tests"}), readability (e.g., \textit{``Nit: indent."}), or overall design, even when the test code is already functional. These suggestions often address issues such as test coverage gaps, naming conventions, redundant logic, or stylistic inconsistencies, aiming to refine the clarity, maintainability, or inline comment quality of the test code.

\textbf{Understanding comments (20.4\%):} On the Gerrit platform, 32\% of test related comments belong to the understanding category. Such comments often reflect the reviewer's doubts or uncertainties about the intent or logic of code modifications. Comments often take the form of requesting clarification (e.g., \textit{``Will there be multiple finish files in this directory at the same time?"} and \textit{``How are those cases different?"}) or expressing suggestions (e.g., \textit{``Do you think we could we make this more specific? maybe have the check be EXPECT\_EQ based on what we expect from the simple mock model?"}). Although these comments may not necessarily constitute explicit modification suggestions, they reflect a lack of confidence in the reviewer's understanding of the code, often requiring the author to provide more context or explanation. However, some comments that end with a question mark are classified as code improvement rather than understanding because they do not express actual uncertainty. Instead, reviewers often use a softer tone or polite phrasing as a form of friendly communication (e.g., \textit{``Could you wrap the followings (line 168$\sim$177) with something like `runBenchmark(``Join Benchmark'')'?''}). These observations highlight the disparate review practices across platforms: Gerrit's stringent environment encourages more intensive scrutiny of test code, whereas GitHub reviewers show a diminished proportional focus on this area.

\textbf{Social Communication (8.6\%):} On the Gerrit platform, this category accounts for 11\% of the total comments. It typically includes remarks such as “LGTM” or “+1”, which are commonly used to express approval. These comments are generally considered social interactions or expressions of sentiment rather than technical feedback, reflecting interpersonal communication between reviewers and authors.

\textbf{Defect (5.5\%):} On the Gerrit platform, 9\% of test-related comments fall into the defect category. This category includes cases where reviewers explicitly raise concerns that the test code may fail during execution. Unlike categories addressing omissions or improvements, defect-related comments focus on incorrect logic, flawed test structure, or invalid assertions that could lead to runtime errors or false test results. These comments are typically more concrete and technical, directly pointing out mistakes that undermine the reliability of the test (e.g., \textit{``This got flagged by internal tests: in graph mode, attempting to `bool()' cast a Tensor raises a specific error instead (`OperatorNotAllowedInGraphError', a subclass of `TypeError'). I think the easiest way is to test that either ValueError os TypeError is raised."}).

\textbf{Knowledge Transfer (4.2\%):} On the Gerrit platform, this category accounts for 4\% of the total comments. It typically includes instances in which the reviewer explicitly aims to share knowledge with the PR author, either by explaining certain concepts or by directing the author to external or internal resources to improve the code. This category includes four subtypes: link, other class, other patch, and example. Among them, 62.5\% comments in this category take the form of examples provided by the reviewer (e.g., \textit{``Need to replace the reference parameter with a pointer parameter. See style guide on reference parameter https://google.github.io/styleguide/cppguide.html"}).

\textbf{Misc (6.6\%):} This category typically consists of comments made at the PR level, without any discussion or suggestions to the test code (e.g., \textit{``Can you add the issue number"}).




\subsubsection{\textbf{GHA Influence: Topic Distribution Perspective}}

When analyzing the influence of GHA on the topics of the reviews, we observed a downward trend in reviewers' discussions regarding improvements to the test code and defect finding. Among the 770 comments analyzed before and after GHA adoption, 51.4\% of the discussions on code improvements occurred before GHA, compared to 48.6\% afterward. Similarly, for discussions concerning defects in test code, 66.7\% occurred before GHA and 33.3\% after. Although these changes are relatively small, they suggest that reviewers may rely more on GHA's automation and, consequently, reduce direct discussions about test code improvements and defect identification.

The Chi-square test yielded a statistic of $\chi^2(5) = 7.53$ with a $p\_value$ of 0.184, which does not meet the threshold for statistical significance ($\alpha = 0.05$). This indicates that there is no significant difference in the distribution of comment categories between the ``Before'' and ``After'' periods. 

\begin{tcolorbox}[colback=black!5,
                  colframe=black,
                  rounded corners=southwest,
                  rounded corners=northwest,
                  boxrule=0.5pt,
                  arc=8pt,
                  left=5pt,
                  right=5pt,
                  top=5pt,
                  bottom=5pt,
                  fonttitle=\bfseries,
                  before skip=10pt,
                  after skip=10pt,
                  breakable]

\textbf{Finding 4.} \textit{Over half of the comments focus on improving already functional test code. Compared to Gerrit, GitHub reviewers appear less concerned with defect detection or code comprehension, as evidenced by the lower proportion of comments in these categories. On GitHub, after GHA adoption, improvement and defect discussions show a slight decline, consistent with greater reliance on automation, despite no significant shift in overall category distributions.}
\end{tcolorbox}

\textbf{Actionable implications:} The frequency of improvement-oriented comments in GitHub test code reviews suggests that reviewers frequently treat test code as a candidate for refinement rather than as a potential source of faults. While such feedback can improve readability and maintainability, it also risks normalizing a ``tests already work'' assumption, thereby reducing attention to defect detection and deep comprehension of test intent. For practitioners, this implies a need to recalibrate review practices on GitHub: teams may benefit from explicitly encouraging reviewers to verify test correctness, adequacy, and failure modes (even when CI passes), rather than limiting feedback to stylistic or refactoring suggestions. Concretely, review guidelines or PR templates could include prompts asking reviewers to assess whether tests would fail under realistic fault scenarios or whether assertions genuinely capture the intended behavior. For tool designers, the findings highlight an opportunity to support deeper test reviews by surfacing signals that CI does not cover (e.g., weak assertions, redundant tests, or untested edge cases), nudging reviewers beyond superficial improvements. Finally, for researchers, the shift away from defect- and comprehension-focused comments underscores the importance of not equating comment volume or activity with review rigor, particularly in PR-based workflows where automation may implicitly discourage critical scrutiny of test logic.

\subsection{RQ3: Do reviewers examine test code before production code during code review, and does GHA adoption alter this order?}

\begin{figure*}[!tp]
    \centering
    \includegraphics[width=1\textwidth]{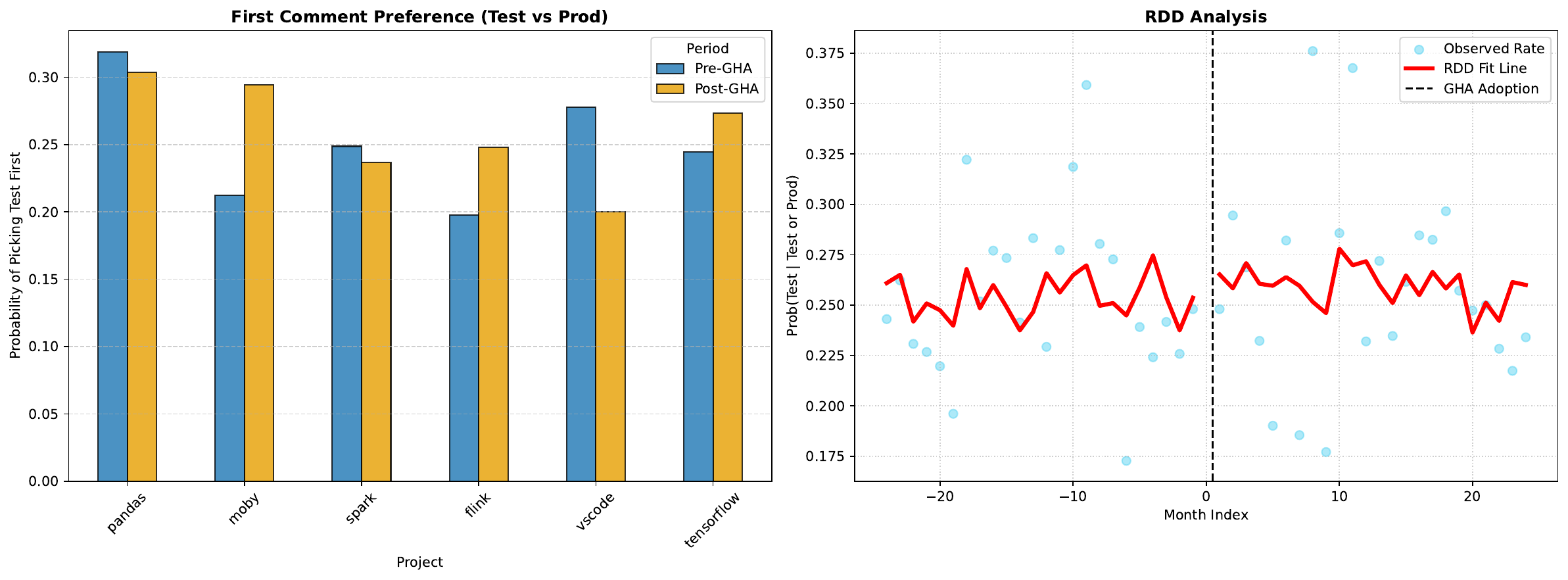}
    \caption{GHA influence on Reviewer Review Preference}
\vspace{-5mm}
    \label{fig:gha_test_review_ratio_comparison_by_project}
\end{figure*}

Figure~\ref{fig:gha_test_review_ratio_comparison_by_project} presents the GHA influence on reviewer preference
across all analyzed projects. We found that only 25.67\% of reviewers commented on test files first, while the majority (74.33\%) reviewed production code first. This finding is consistent with the results of the \textit{original study}~\citep{spadini2018testing} and with common expectations, as reviewers typically devote their limited time to areas perceived as more critical, such as production code. Our quantitative analysis complements the \textit{original study}.

We find no statistically significant evidence that the introduction of GHA altered reviewers' initial focus toward test code, as measured by whether the first review comment was posted on a test file based on timestamps.
Across the six projects, the probability that the first review comment targeted test files remained remarkably stable, increasing only marginally from 25.36\% (Pre-GHA, N = 3,226) to 26.02\% (Post-GHA, N = 2,940).
Regression discontinuity analysis further corroborates this stability: the estimated coefficient for the GHA adoption indicator (PostGHA) is not statistically significant (p = 0.862), providing no evidence of an immediate or discontinuous shift in review behavior at the adoption cutoff.
In contrast, production churn exhibits a highly significant association with reviewers’ initial focus ($p=0.004$). This is highly intuitive, as extensive modifications to production logic naturally draw reviewers' primary attention, potentially overshadowing the accompanying test code.

While knowing whether a reviewer comments first on test files or production files does not, by itself, directly establish a causal link to review quality, our intention is not to claim such causality but to document a measurable reviewer preference that can be consistently extracted across projects. More specifically, our goal is to complement the interview-based findings of the \textit{original study}~\citep{spadini2018testing} by providing a large-scale, data-driven perspective. This allows us to examine whether the observed reviewer behaviors remain consistent or undergo structural shifts when tested against a broader empirical dataset under the influence of GHA adoption.

Our dataset records, for every PR, whether reviewers commented first on the test code or the production code. This provides a valuable statistical signal that can serve as a foundation for future research. For example, subsequent studies should investigate whether PRs where reviewers initially focused on test code tend to be more stable or less bug-prone over time, or whether different review orders correlate with distinct review models that impact project outcomes. We believe our contribution lies in surfacing and quantifying this behavioral pattern, thereby enabling the community to further explore its long-term implications for review quality and system reliability.

\begin{tcolorbox}[colback=black!5,
                  colframe=black,
                  rounded corners=southwest,
                  rounded corners=northwest,
                  boxrule=0.5pt,
                  arc=8pt,
                  left=5pt,
                  right=5pt,
                  top=5pt,
                  bottom=5pt,
                  fonttitle=\bfseries,
                  before skip=10pt,
                  after skip=10pt,
                  breakable]

\textbf{Finding 5.} \textit{Adopting GHA does not change reviewers' initial focus on production code. Instead, their attention remains stable and is primarily driven by the amount of production code churn in a pull request.}
\end{tcolorbox}

\textbf{Actionable Implications:} The persistence of a production-first strategy despite GHA adoption suggests that automation alone is insufficient to shift reviewers’ cognitive priorities. For practitioners, this implies that CI tooling must be supplemented by explicit interventions, such as PR templates or decoupled review passes, to ensure test scrutiny in high-churn PRs. For tool designers, this highlights an opportunity for ``attention-aware'' interfaces that prioritize test visibility when production changes are extensive. Finally, for researchers, the dominant effect of production churn underscores the need to control for change magnitude in review studies; failing to do so may confound tool-driven effects with ingrained, workload-driven reviewer heuristics.


\section{Discussion}\label{sec5}

This study presents a replication and extension of the \textit{original study}~\citep{spadini2018testing} by examining a broader range of projects on GitHub, the most representative platform of the PR review model. 
The results for RQ1 confirm that reviewers primarily focus on production code, though this tendency is somewhat mitigated on GitHub compared to other platforms. We further demonstrate that differences in review models inherently dictate variations in review density and argue that aggregating project data can mask significant internal disparities in review practices. While the adoption of GHA triggered brief, sharp shifts in review behavior, a sustained long-term decline in test review was observed in projects such as Pandas and Spark. 
For RQ2, our qualitative analysis reveals that GitHub reviewers often remain at a superficial level, overlooking deep-seated logic issues in test code. Finally, addressing 
RQ3, we identify production code churn as a decisive factor influencing a reviewer's initial focus; specifically, larger churn significantly decreases the likelihood of an early test case review.

\newcolumntype{I}{>{\hsize=0.75\hsize\raggedright\arraybackslash}X} 
\newcolumntype{M}{>{\hsize=1.25\hsize\raggedright\arraybackslash}X}

\begin{sidewaystable*}[!tp]
\footnotesize 
\centering
\caption{Insights and Implications Across Stakeholders from This Study}
\label{tab:combined_implications_grouped}

\begin{tabularx}{\textheight}{
  l 
  >{\raggedright\arraybackslash}p{6em} 
  I 
  M 
  >{\raggedright\arraybackslash}p{5em}
}
\toprule
\textbf{Type} & \textbf{Dimension} & \textbf{Key Insight} & \textbf{Implication} & \textbf{SH} \\ \midrule

\multirow{9}{*}{\rotatebox[origin=c]{90}{Platform}} & 
\textbf{Platform Culture} & 
Gerrit fosters a stricter, correctness-oriented review culture, whereas GitHub encourages a more flexible and lightweight style. & 
\textbf{TLs} on GitHub platform should not rely on the platform's default mechanisms but introduce tools to increase the likelihood that code is reviewed. \textbf{Rs} need to recognize how platform differences may influence experimental outcomes in the future research. Specifically, Gerrit yields denser reviews, whereas GitHub provides a more balanced focus on production and test code. & 
TL, R \\ \cmidrule{2-5}

& \textbf{Test Code Dist.} & 
In GitHub PRs, reviewers tend to focus on suggesting improvements to test code, whereas on Gerrit, reviewers are more inclined to identify defects in test code and to express a lack of understanding of specific test code segments. & 
\textbf{TLs} should recognize that while GitHub reviewers tend to focus on surface-level suggestions for test code, Gerrit reviewers are more inclined to pursue deeper understanding and defect detection. On GitHub, \textbf{REs} should complement suggestion-oriented feedback with stricter defect detection to avoid overlooking risks. \textbf{Rs} may explore the organizational and cultural factors that shape the differing distributions of test code review topics across platforms. & 
TL, RE, R \\ \midrule

\multirow{4}{*}{\rotatebox[origin=c]{90}{Project}} & 
\textbf{Project Variability} & 
Review practices vary substantially across projects on the same platform (e.g., Moby vs. VSCode). & 
\textbf{Rs} should recognize that projects differ in the level of attention devoted to test code during review. Identifying the factors that contribute to such differences, as well as their long-term consequences, remains an important avenue for future work. \textbf{TLs} should be aware that in projects where system correctness is critical, test code warrants explicit emphasis, and reviewers may need targeted guidance to prioritize it, especially when integrating with current LLM generated test code~\citep{haider2026understanding, cynthia2026we} where reviewers may lack the cognitive bandwidth to scrutinize them. & 
R, TL \\ \midrule

\multirow{10}{*}{\rotatebox[origin=c]{90}{Review Dynamics}} & 
\textbf{GHA Short-term Influence} & 
GHA triggers a temporary reallocation of review effort toward test suites, followed by a subsequent pivot back to production code. & 
Our study further indicates that \textbf{Rs} should realize the surge in test review effort is merely a transient phenomenon tied to the migration phase. For \textbf{tool designers}, this finding highlights an opportunity that persistently surfaces test quality signals at review time could counterbalance production-centric attention and better align reviewer effort with long-term quality risks which will cause more resources waste~\citep{turzo2025first, bouzenia2024resource, labuschagne2017measuring}.& 
R, TD \\ \cmidrule{2-5}

& \textbf{GHA Long-term Influence} & 
GHA adoption reduces both the review rate and density of test code in a long time, fostered a strong sense of automation trust, sometimes to the point of overreliance, leading to reduced manual inspection and maintenance of test code on some projects. & 
Reduced engagement with test code may accumulate technical debt. \textbf{Rs} could conduct longitudinal studies to quantify CI’s long-term impact on test quality and maintenance costs. Prior work has shown that CI/DI usage can involve substantial waste \citep{bouzenia2024resource, labuschagne2017measuring}. \textbf{Rs} in CI/DI efficiency should also recognize that neglecting test code during review may lead to duplicated or low-quality tests, increased GHA failures, and ultimately reduced efficiency. \textbf{TLs} should take into account the long-term impact of CI and update project review guidelines to enhance the effectiveness of test code reviews~\citep{turzo2024makes}. & 
R, TL \\ \cmidrule{2-5}

& \textbf{Attention Order} & 
Reviewers prioritize production code first on both platforms; reviewers are more prone to neglecting test code when a PR contains substantial production code churn.& 
Prior work~\citep{fregnan2022first} shows that file order can influence review outcomes. \textbf{Rs} could further examine the long-term impact of test-versus-production review order on project quality. \textbf{TDs} could decouple test and production reviews when manage a project. \textbf{REs} should recognize this phenomenon and adopt strategies to mitigate cognitive load, ensuring that test code receives adequate attention.& 
R, TD, RE \\ \bottomrule

\addlinespace[1ex]
\multicolumn{5}{l}{\scriptsize \textbf{SH (Stakeholders):} TL: Team Leads; R: Researchers; RE: Reviewers; TD: Tool Designers.} \\
\end{tabularx}
\end{sidewaystable*}

Table~\ref{tab:combined_implications_grouped} synthesizes the key insights from our empirical findings and distills their implications across multiple stakeholder groups, including team leads, reviewers, tool designers, and researchers. The table's content is organized along the platform, project, and review-dynamics dimensions. Furthermore, the table highlights how review culture, platform properties, and CI automation shape attention to test code, both in the short and long term.

\section{Threats to Validity}
Every empirical study is subject to limitations that can affect the interpretation and generalizability of its findings. This section acknowledges and discusses potential threats to the validity of our study, which are categorized into four main types~\citep{Wohlin2000}: internal, construct, conclusion, and external validity.

\paragraph{Internal Validity}\label{sec6}
We used a time-based RDD to examine whether adopting GHA changed reviewer behavior. A potential threat to internal validity arises from substantial project-level heterogeneity in review practices. Our design assumes that, absent GHA adoption, projects would exhibit comparable temporal evolution in reviewers’ relative attention to production versus test code. However, we observe that certain projects deviate markedly from this pattern. In particular, the project fixed effect for VSCode is large and statistically significant, indicating a substantially higher baseline log odds ratio compared to other projects. This persistent deviation suggests that VSCode follows a distinct review regime, potentially driven by factors such as program language or contributor structure. Pooling such a project with others may weaken the plausibility of the parallel trends intuition underlying our time-based regression discontinuity design. To address this concern, we conduct a sensitivity analysis by re-estimating our models after excluding VSCode. Importantly, we report results from both the full sample and the restricted sample to ensure transparency. While the full-sample analysis shows no statistically significant discontinuity, the restricted model reveals a significant immediate shift at the GHA adoption cutoff. This comparison suggests that heterogeneity across projects can mask discontinuities among more comparable units. We emphasize that this exclusion is not motivated by statistical significance alone, but by methodological considerations related to identification assumptions. 

\paragraph{Construct Validity} To replicate the experimental results using the same approach as the \textit{original study}, we rely on file paths and naming conventions to distinguish between test, production, and configuration files. This heuristic method may lead to misclassification, potentially affecting the accuracy of our conclusions. However, during the hybrid card sorting for RQ2, we confirmed the validity of our naming convention method, thereby reducing the risk posed by this threat.
Another threat is related to annotation during card sorting, which is inherently subject to raters’ subjective judgment. 
To mitigate this, we conducted six iterative coding rounds to improve inter-rater agreement. Nevertheless, residual interpretation bias may still affect the accuracy of our analysis.

\paragraph{Conclusion Validity}
To ensure the reliability of our statistical inferences, we conducted diagnostic checks for all regression models, including tests for residual normality and homoscedasticity. Since our review metrics were highly skewed, we employed non-parametric tests and median-based indicators to mitigate the risk of violated assumptions. We also reported regression coefficients and Odds Ratios to provide a clear measure of effect magnitude beyond mere p-values. Furthermore, to address the risk of subjective bias in qualitative analysis, two authors conducted multiple iterative coding rounds until achieving high inter-rater agreement (77\%), thereby ensuring the robustness of the categories identified in RQ2.

\paragraph{External Validity}  
To ensure rich review data, we selected large, representative open-source projects with high activity. We know that review practices in smaller-scale or commercial/private projects may differ, and our findings may not generalize to those settings.
Moreover, one key part of our study revisits findings from an earlier Gerrit-based investigation conducted in 2018. We acknowledge that both ecosystems have evolved since then (e.g., with the integration of new tools such as GHA and LLMs), and the contexts are not perfectly equivalent. However, this difference is central to the contribution of a replication study: rather than assuming direct comparability, we document consistencies and divergences across time and platforms. This allows the community to see which observations appear robust and enduring, and which are platform- or era-specific.

\section{Related Work}
The \textit{Original study}~\citep{spadini2018testing} has emphasized that test code is as important as production code, with a comparable likelihood of introducing future defects \citep{spadini2018testing, esfandyaridoulabi2025adaptive}.  Moreover, other studies have shown that the quality of test code directly affects the quality of production code~\citep{athanasiou2014test, zaidman2008mining}. Notably, Vahabzadeh et al.~\citep{vahabzadeh2015empirical} reported that nearly half (47\%) of the projects they examined contained bugs in test code that were later reported and fixed. In addition, review order has been shown to influence the final quality and outcomes of code reviews \citep{fregnan2022first}. Collectively, these findings highlight the critical importance of studying test code reviews alongside production code reviews.

Code review has undergone a long period of development and has been proven effective in gate-keeping and improving code quality~\citep{badampudi2023modern}. Code review efficiency has also attracted scholarly attention~\citep{turzo2024makes, bosu2015characteristics}, as even marginal improvements can translate into substantial time and cost savings for development teams. However, most of the existing studies focus on production code~\citep{bacchelli2013expectations, sadowski2018modern, mcintosh2016empirical}, while the review of test code, despite being equally important, has often been overlooked.
Aniche et al.~\citep{}{aniche2021developers} found six core elements, such as documentation, source code, and mental models, important for developers designing test cases. They also concluded that developers often rely on documentation to build their initial mental models, and continually validate or revise their understanding by reading source code and running test code. Furthermore, developers use branch coverage as a basis for stopping testing. Despite its relevance, the study by Aniche et al.~\citep{aniche2021developers} focuses on the test creation phase and has not examined the review practice of the test code during the code review process. Test-driven code review was investigated in a controlled experiment and shows that this practice does not significantly affect the overall quality and quantity of review comments~\citep{spadini2019test}. This process is fundamental to finding test-related issues.

CI links test and production code together to support both the software development and code review process~\citep{fowler2006continuous, duvall2007continuous}. Consequently, the CI influence on code quality has been studied~\citep{soares2022effects}. Previous studies~\citep{zhao2017impact, nery2019empirical, zampetti2019study} investigated the various impacts that Travis CI and GHA have on test ratio, test coverage, and PR status (e.g., whether a PR is eventually merged or closed, affected by GitHub CI results). Rahman and Roy~\citep{rahman2017impact} show that projects with frequent CI builds tend to maintain a steady and higher quality level of code review activity over time. Golzadeh et al.~\citep{golzadeh2022rise} show that within only 18 months of its release, GHA surpassed Travis CI, 
with over 3,700 repositories (37.5\% of all migrations from Travis to GHA since GHA) migrated to GHA within just three months.

While there are a few studies focused on test code review within Gerrit~\citep{spadini2018testing, spadini2019test}, they overlook platform and project differences. Besides, the most popular CI tool, GHA, was introduced in 2019 and has had a sustained impact on reviewers' practice since then. To address this gap, our work combines qualitative and quantitative methods to investigate how platform differences and GHA adoption have influenced the review of test code.

\section{Conclusion}

In this work, we replicated Spadini et al.’s study~\citep{spadini2018testing}. While they study Gerrit, our focus is on GitHub PRs to examine reviewers’ attention to test code, with particular attention to the influence of GHA. 
Our results demonstrate that differences in review models inherently dictate variations in review density; specifically, the strict pre-commit model of Gerrit increases the likelihood of comments per file compared with GitHub's model. We also reveal a critical methodological insight: aggregating data at the platform level only provides a partial view of review practices. In fact, project-level heterogeneity can lead to conflicting practices that cancel each other out, thereby confounding platform-wide analysis.

Through a granular project-level investigation, we found that upon excluding VSCode, GHA adoption led to a brief, high-intensity surge in test code review, which subsequently shifted toward production code. While GHA's impact on platform-level test review practices was limited, we observed a significant, sustained long-term decline in both the probability and frequency of test code reviews across major projects such as Pandas and Spark. Our qualitative coding analysis further illuminates this shift: GitHub reviewers tend to focus on superficial improvements such as formatting and naming, whereas Gerrit reviewers are more likely to scrutinize the underlying logic and bugs within test code. This tendency toward superficiality showed a non-significant but observable trend of intensification following GHA adoption.
Quantitatively, our study shows that 74.33\% of reviewers’ initial focus is on production code. This bias is not mitigated by GHA but is instead strongly conditioned by production code churn; specifically, higher churn makes reviewers more likely to overlook test code. To support community efforts, all experimental data, scripts, and results are provided in a replication package, with much of the code reusable for GitHub repository mining research.

The contributions of our work have implications for diverse stakeholders. For researchers, we underscore the need to account for platform- and project-level variance. Our work provides a baseline for distinguishing the impacts of GHA from those of AI-driven automation tools, both of which aim to accelerate review efficiency. For reviewers and team leads, we emphasize that explicit guidance for test code review remains paramount. Given the unknown long-term impact of the synergy between automated verification and LLMs on project quality, reviewers must strive for deep logical understanding rather than superficial styling fixes. Finally, for tool designers, our results suggest that providing more detailed feedback on test execution traces and performance would be a meaningful intervention to re-balance reviewer attention~\citep{guan2025faster}. This study demonstrates the feasibility of future work on the qualitative factors underlying project heterogeneity and the relationship between reviewers' initial focus and long-term code quality.

Building on our replication study, future research should investigate the downstream consequences of reduced test code review on software quality, including test fragility, test debt, and defect leakage over longer time periods. Longitudinal studies could examine whether reduced reviewer engagement with tests under CI automation (e.g., GHA) leads to measurable degradation in test effectiveness or maintainability. Additionally, future work should explore interventions (e.g., review tooling, workflow nudges, or policy changes) that explicitly rebalance reviewer attention toward test code without negating the efficiency benefits of automation. Finally, as LLM-based tools increasingly generate tests, studies can be performed to disentangle the effects of CI automation from AI-assisted development and to understand how reviewers evaluate machine-generated test code in practice. Future research could utilize our categories in RQ2 to investigate whether humans review AI-generated tests with the same rigor, or if the automation trust we observed with GHA leads to even greater marginalization of test quality when AI is the primary author.









\bibliography{sn-bibliography}

\end{document}